\documentclass[aps,pre,epsfigure,twocolumn,superscriptaddress]{revtex4-1}
\usepackage[colorlinks=true,linkcolor=blue,urlcolor=blue,citecolor=blue,pdfusetitle]{hyperref}
\usepackage[utf8]{inputenc}
\usepackage[english]{babel}
\usepackage{amsmath}
\usepackage[caption = false]{subfig}
\usepackage{graphicx,epstopdf}
\usepackage{blindtext}
\usepackage{lipsum}
\usepackage{amsfonts}
\usepackage{bbm}
\usepackage{amssymb}
\usepackage{enumerate}
\usepackage{color}
\usepackage{latexsym}
\usepackage{times,txfonts}

\newcommand{\Hcal}{\mathcal{H}}

\newcommand{\Pcal}{\mathcal{P}}
\newcommand{\Acal}{\mathcal{A}}

\newcommand{\Ecal}{\mathcal{E}}

\newcommand{\Zcal}{\mathcal{Z}}

\newcommand{\Ical}{\mathcal{I}}

\newcommand{\Ccal}{\mathcal{C}}
\newcommand{\Bcal}{\mathcal{B}}
\newcommand{\Ucal}{\mathcal{U}}

\newcommand{\1}{\mathbbm{1}}

\newcommand{\Rmath}{\mathbbm{R}}

\newcommand{\ket}[1]{| #1 \rangle}

\newcommand{\bra}[1]{\langle #1 |}

\newcommand{\tr}[1]{ \text{Tr}\left\{ #1 \right\}}
\newcommand{\trs}[1]{ \text{Tr} \{ #1 \}}

\begin{document}

\title{Stable and charge-switchable quantum batteries}

\author{Alan C. Santos}
\email{ac\_santos@id.uff.br}
\affiliation{Instituto de F\'{i}sica, Universidade Federal Fluminense, Av. Gal. Milton Tavares de Souza s/n, Gragoat\'{a}, 24210-346 Niter\'{o}i, Rio de Janeiro, Brazil}

\author{Andreia Saguia}
\email{amen@if.uff.br}
\affiliation{Instituto de F\'{i}sica, Universidade Federal Fluminense, Av. Gal. Milton Tavares de Souza s/n, Gragoat\'{a}, 24210-346 Niter\'{o}i, Rio de Janeiro, Brazil}

\author{Marcelo S. Sarandy}
\email{msarandy@id.uff.br}
\affiliation{Instituto de F\'{i}sica, Universidade Federal Fluminense, Av. Gal. Milton Tavares de Souza s/n, Gragoat\'{a}, 24210-346 Niter\'{o}i, Rio de Janeiro, Brazil}

\begin{abstract}
A fully operational loss-free quantum battery requires an inherent control over the energy transfer process, with the ability of keeping the energy retained with no leakage. Moreover, it also requires a stable discharge mechanism, which entails that no energy revivals occur as the device starts its energy distribution. Here, we provide a scalable solution for both requirements. To this aim, we propose a general design for a quantum battery based on an {\it{energy current}} (EC) observable quantifying the energy transfer rate to a consumption hub. More specifically, we introduce an instantaneous EC operator describing the energy transfer process driven by an arbitrary interaction Hamiltonian. The EC observable is shown to be the root for two main applications: (i) a trapping energy mechanism based on a common eigenstate between the EC operator and the interaction Hamiltonian, in which the battery can indefinitely retain its energy even if it is coupled to the consumption hub; (ii) an asymptotically stable discharge mechanism, which is achieved through an adiabatic evolution eventually yielding vanishing EC. These two independent but complementary applications are illustrated in quantum spin chains, where the trapping energy control is realized through Bell pairwise entanglement and the stability arises as a general consequence of the adiabatic spin dynamics. 
\end{abstract}

\maketitle

\section{Introduction}

Exploring inherently quantum features of physical systems in order to realize high performance tasks has been the ultimate ideal pursued by 
quantum technologies. As an illustration, entanglement, which is a genuinely quantum resource, has been employed as a key concept to design quantum devices, 
such as quantum transistors~\cite{Loft:18,Marchukov:16,dePonte:19}, quantum heat engines~\cite{kieu:04,Ueda:13,Lindenfels:19}, image 
processors~\cite{Abouraddy:02,Gatti:04,Valencia:05,Bornman:19}, among 
others~\cite{Giovannetti:11,Nielsen:Book,Deffner-Campbel:Book}. 
In this scenario, a specific quantum device that may strongly benefit from entanglement is a {\textit{quantum battery}} (QB)~\cite{Alicki:13,PRL2013Huber,PRL2017Binder,PRL_Andolina}, 
which is a quantum system able to both temporarily store and then transfer energy. QBs are potentially  
relevant as fuel for other quantum devices and, more generally, for boosting a proper development of quantum networks. 
In particular, they have been proposed in a number of distinct 
experimental architectures, such as spin systems~\cite{Le:18}, quantum cavities~\cite{Binder:15,Fusco:16,Zhang:18,Ferraro:18}, superconducting transmon qubits~\cite{Santos:19-a}, quantum oscillators~\cite{Andolina:18,Andolina:19} and spin batteries in quantum dots~\cite{Long:03}. 

Among fundamental challenges for useful QBs are both the control of the energy transfer and the stability of the discharge process to an available consumption hub (CH) 
(see, e.g., Refs.~\cite{Santos:19-a,Gherardini:19,Kamin:19} for recent discussion about the stability topic).    
In this work, we aim at providing a general approach to solve each of these problems. More specifically, we will focus on situations where the battery is initially charged and 
an optimal energy transfer process driven by an interaction Hamltonian for the QB and the CH is desired. This will be achieved by introducing a quantum physical 
observable that describes the instantaneous energy transfer rate to the CH, which will be referred to as the \textit{energy current (EC) operator}. By exploring its quantum properties, 
we will show that the EC operator is the root for two main applications: (i) a trapping energy mechanism based on a common eigenstate with the interaction Hamiltonian, 
in which the battery can indefinitely retain its energy even if it is coupled to the CH; (ii) an asymptotically stable discharge mechanism, which will be achieved through an 
adiabatic evolution leading to vanishing EC. These two applications will then be illustrated in quantum spin chains. For the trapping energy mechanism, 
quantum control will be provided by pairwise entanglement through Bell quantum states for the spins within the battery cell. Concerning the stability, it will be shown to arise 
as a general consequence of a suitably arranged adiabatic spin dynamics. 

\section{Quantum batteries and consumption hubs}

Consider a composite quantum system described by a Hilbert space $\Bcal \otimes \Acal$, where $\Bcal$ is associated with a QB and $\Acal$ refers to an auxiliary system, which will play the role of a CH. The composite system is driven by the Hamiltonian $H(t)\!=\!H_{0} + H_{\text{C}}(t)$, where $H_{0}$ is the individual energy contribution for both QB and CH subsystems, while $H_{\text{C}}(t)$ is the corresponding charging Hamiltonian. The Hamiltonian $H_{0}$ can be written as $H_{0}\!=\!H^{\Bcal}_{0} + H^{\Acal}_{0}$, with $H^{\Bcal}_{0}$ being the battery inner Hamiltonian and $H^{\Acal}_{0}$ the auxiliary inner Hamiltonian. In general, the inner parts of the QB may interact with each other, with this inner interaction included in Hamiltonian $H^{\Bcal}_{0}$.

The success of the energy transfer from the QB to the CH can be measured by the {\textit{ergotropy}, i.e. the maximum amount of work which can be further extracted from the QB. The ergotropy of a quantum system can be generically defined by
\begin{align}
\Ecal = \max_{V\in \Ucal} \left( \tr{H\rho} - \trs{HV\rho V^\dagger}\right) \text{ , }
\end{align}
where $H$ is the reference Hamiltonian of the system and $\Ucal$ is the set of unitary operations~\cite{Perarnau-Llobe:15}. The maximization over all $V \in \Ucal$ is associated with the {\it passive state} $\sigma_{\rho}$, i.e., $\sigma_{\rho}$ is the state in which no energy can be extracted from the system by unitary operations~\cite{Allahverdyan:04}. In general, there are a number of possibilities for a passive state~\cite{Perarnau-Llobe:15}, which are related to each other by unitary rotations in the degenerate subspaces of the reference Hamiltonian. In cases where we are dealing with discharging processes of QBs from a pure state, we identify the ergotropy simply as the difference of energy from the initial state and the ground state of the system. In fact, since we can find a operator $V_{\psi}$ so that we can drive the system from $\rho_{\psi}\!=\!\ket{\psi}\bra{\psi}$ to the ground state $\rho_{g}\!=\!\ket{g}\bra{g}$, then we can write $\Ecal\!=\!\trs{H\rho_{\psi}} - \trs{H \rho_{g}}$. In particular, $\rho_{g}$ is identified as a passive state and we write its corresponding energy as	$E^{\Acal}_{\text{emp}}=\trs{H \rho_{g}}$.
	
Concerning the CH, its energy at the initial time  
reads $\Ecal^{\Acal}_{0}\!=\!\tr{\rho_{0}^{\Acal}H^{\Acal}_{0}} - E^{\Acal}_{\text{emp}}$, where $\rho_{0}^{\Acal}$ is the CH initial (pure) state. 
Notice that, by starting the evolution in the ground state of the CH, we have $\Ecal^{\Acal}_{0}=0$. In this paper, we are interested in an energy transfer process from a QB (initially charged in a pure state) to the CH. 
We can do this by coupling the QB to the CH, an energy transfer process will take place,  
which will be governed by Schr\"odinger equation. It is useful to adopt the interaction picture, where the composite density operator $\rho_{\text{int}}(t)$ is driven by 
$i\hbar\dot{\rho}_{\text{int}}(t)\!=\![H_{\text{int}}(t),\rho_{\text{int}}(t)]$, with the dot symbol denoting time derivative and $H_{\text{int}}(t)$ the new charging Hamiltonian in the picture interaction. 
The amount of transferred energy $C(t)$ at instant $t$ is given by the instantaneous energy variation $U(t) - U(0)$ 
in the CH, reading 
\begin{eqnarray}
C(t)  = U(t) - U(0) = \tr{H_{0}^{\Acal}\rho_{\text{int}}(t)} - E^{\Acal}_{\text{emp}} \text{ . }
\label{Ct}
\end{eqnarray}
Eq.~(\ref{Ct}) provides the amount of energy transferred from the QB to the CH.
Notice that, if the reduced density operator of the CH corresponds to a pure state at the end of the transfer process, at $t=\tau$,
then it is straightforward to show that $C(\tau)$ can be obtained from the CH ergotropy at the end of the evolution. In this case, since 
its ergotropy is nonvanishing, it is possible to distribute energy from the CH to other systems.

\section{The energy current operator}

The starting point for the QB proposal is to consider the instantaneous EC, which is defined as $\Pcal(t)\!=\!\dot{C}(t)$. Notice that $\Pcal(t)$ 
quantifies the rate of internal energy transferred to the CH. 
It is worth mentioning that $\Pcal(t)$ does not necessarily quantify the instantaneous rate 
of ergotropy, since the CH may evolve to mixed states at intermediate times. However, considering the initial and final states as pure states, the EC
will coincide with the power in the sense of time derivative of work.
We can show that $\Pcal(t)$ can be obtained as the expectation value of a Hermitian EC operator $\hat{\Pcal}(t)$, 
namely $\Pcal(t)\!=\!\trs{\hat{\Pcal}(t) \rho_{\text{int}}(t)}$. More specifically, we can obtain $\hat{\Pcal}(t)$ in terms of the Hamitonian of the system as (see Appendix~\ref{ApA})
\begin{eqnarray}
\hat{\Pcal}(t) = (1/i\hbar)[H_{0}^{\Acal},H_{\text{int}}(t)] \text{ . } \label{PowerOperator}
\end{eqnarray}

\subsection{Energy trapping mechanism} 

As a first application of the observable $\hat{\Pcal}(t)$, we introduce a general {\it energy trapping mechanism}. To begin with, assume a constant Hamiltonian $H_{\text{int}}$,
which implies that $\hat{\Pcal}(t)\!=\!\hat{\Pcal}$ is also a constant operator. Then, notice that, if $\hat{\Pcal}$ and $H_{\text{int}}$ commute, 
we have that  $\hat{\Pcal}$ is a constant of motion. Assume now that there is a common eigenstate $\ket{p_{0}}$ of $\hat{\Pcal}$ and $H_{\text{int}}$ with EC eigenvalue $p_{0}\!=\!0$. 
Then, by initially preparing the quantum system at the quantum state $\ket{p_{0}}$, no amount of energy can be extracted from the QB. In fact, let us consider the 
initial state of the system as $\ket{p_{0}}$, where we have $\hat{\Pcal} \ket{p_{0}}\!=\!0$ and $H_{\text{int}}\ket{p_{0}}\!=\!E_{p0}\ket{p_{0}}$. The evolved state reads 
$\ket{\psi(t)}\!=\!e^{-iE_{p0}t/\hbar}\ket{p_{0}}$. Then, the instantaneous EC is $\Pcal(t)\!=\!\bra{\psi(t)} \hat{\Pcal} \ket{\psi(t)}\!=\!\bra{p_{0}} \hat{\Pcal} \ket{p_{0}}\!=\!0$ 
and no amount of energy can be introduced or extracted from the system driven by $H_{\text{int}}$. Remarkably, this conclusion holds even if $\hat{\Pcal}$ and $H_{\text{int}}$ 
does {\it not} commute, as long as they share at least a single eigenstate with EC eigenvalue $p_{0}\!=\!0$. This less restrictive situation is indeed even more interesting and will 
be explored in the next section. Indeed, the trapping mechanism opens perspectives for a class of QBs in which the energy is not transferred even when the battery is connected 
to the CH. 
Notice that, even though energy transfer is inhibited, there may occur charge leakage through coherence transfer~\cite{Hovhannisyan:19,James:20,Kamin:19}. 
Here, the mechanism for energy trapping ensures absence of charge leakage for {\it time-independent} Hamiltonians, since the {\it stationary} dynamics will keep the 
global state fixed (up to a phase) during the quantum evolution. In the more general case of a time-dependent Hamiltonian, a non-stationary 
dynamics could admit energy trapping in average, but not necessarily ergotropy trapping. As we shall see, for time-independent Hamiltonians, 
we can also develop an activation mechanism, where we can control the exact time to discharge the energy to the CH.

\subsection{Stability of quantum batteries through adiabatic dynamics}

As a second application of the observable $\hat{\Pcal}(t)$, we can show that the \textit{adiabatic dynamics} allows for a {\it stable energy transfer to the CH}. We define that the system 
QB-CH undergoes an adiabatic energy transfer process when the composite state $\ket{\psi(t)}$ evolves adiabatically under the Hamiltonian $H(t)$ that drives the system. 
Let the system be initialized in the state $\ket{\psi(0)}\!=\!\sum_{n} c_{n} \ket{E_{n}(0)}$, where $\{\ket{E_{n}(0)}\}$ is the set of instantaneous eigenstates of $H(0)$. 
Then, in the adiabatic regime, we have 
\begin{align}
\ket{\psi_{\text{ad}}(t)}\!=\!\sum_{n} c_{n} e^{i\theta^{\text{ad}}_{n}(t)}\ket{E_{n}(t)} \text{ , }
\end{align}
where $\theta^{\text{ad}}_{n}(t)$ are the adiabatic phases accompanying the adiabatic dynamics~\cite{Berry:84}. In this regime, the expected value of the EC operator is (see Appendix~\ref{ApB})
\begin{eqnarray}
\Pcal_{\text{ad}}(t) = \frac{1}{i\hbar} \sum\nolimits_{n,m} c_{n} c^{\ast}_{m} e^{i\Delta^{\text{ad}}_{nm}(t)} E_{mn}(t)\bra{E_{n}(t)}H_{0}^{\Acal}\ket{E_{m}(t)} \text{ , }
\label{potadi}
\end{eqnarray}
where $\Delta^{\text{ad}}_{nm}(t)\!=\!\theta^{\text{ad}}_{n}(t) - \theta^{\text{ad}}_{m}(t)$ and $\Delta E_{mn}(t)\!=\!E_{m}(t)-E_{n}(t)$. Now, notice that if we start the evolution 
in a single eigenstate (or a set of degenerate eigenstates) of the whole system, then $\Pcal_{\text{ad}}(t)\!=\!0$ because $\Delta \tilde{E}_{mn}(t)\!\rightarrow\!\Delta \tilde{E}_{mm}(t)\!=\!0$. Since the stability condition 
for an interval $\Ical : [t_{1},t_{2}]$ can be mathematically written as $\Pcal_{\text{ad}}(t)\!=\!0$ for any $t \in \Ical$, then Eq.~(\ref{potadi}) can be shown to imply that the 
adiabatic dynamics provides a general stable charging process, since $\Pcal_{\text{ad}}(t\geq \tau_{\text{ad}})\!=\!0$ where $\tau_{\text{ad}}$ is the required time scale to achieve 
the adiabatic regime. Although the condition of a system starting in a single eigenstate is not a necessary condition for stable charging/discharging process, it is important to mention that this is a \textit{sufficient condition} to achieve stability of the battery.

\section{Bell Quantum batteries}

To explore the role of quantum correlations and illustrate the trapping mechanism previously derived, let us consider a CH given by a single qubit and a 
QB composed by a single cell, where the cell contains two non-interacting qubits. Notice that, since there is no interaction inside the QB cell, we do not need to engineer 
any internal coupling in the battery. Therefore, the bare Hamiltonian for each qubit is given by $H_{0}\!=\!\hbar \omega (\ket{1}\bra{1} - \ket{0}\bra{0})$, with $\ket{0}$ and $\ket{1}$ 
denoting the ``empty" and the ``full" charge state, respectively. When the QB is coupled to the CH, the whole system evolves under action of the interaction Hamiltonian
\begin{eqnarray}
H_{\text{C}} = \hbar J \sum\nolimits_{n=1}^{2}\left( \sigma_{x}^{\Bcal_n}\sigma_{x}^{\Acal} + \sigma_{y}^{\Bcal_n}\sigma_{y}^{\Acal} \right) \text{ , } \label{HCharge}
\end{eqnarray}
where $\sigma_{i}^{\Bcal_n}$ ($i=x,y$) is a Pauli operator acting on the Hilbert space $\Hcal_{\Bcal_n}$ of the $n$-th qubit of the QB ($n=1,2$) and 
$\sigma_{i}^{\Acal}$ acts on $\Hcal_{\Acal}$. A schematic representation of this configuration is shown in Fig.~\ref{FigScheme}. 
From Eq.~\eqref{PowerOperator} the EC operator is
\begin{eqnarray}
\hat{\Pcal} = 2J \omega \hbar 
\sum\nolimits_{n=1}^{2}\left( \sigma_{x}^{\Bcal_n}\sigma_{y}^{\Acal} - \sigma_{y}^{\Bcal_n}\sigma_{x}^{\Acal} \right) \text{ , } \label{PowerOperatorSystem}
\end{eqnarray}
where we used that $H_{\text{int}}\!=\!H_{\text{C}}$. Even though $\hat{\Pcal}$ and $H_{\text{int}}$ do {\it not} commute, we identify a quantum state that is a simultaneous eigenstate of both $\hat{\Pcal}$ and $H_{\text{int}}$, which reads ($\ket{\beta_{11}}_{\Bcal}$ denoting a Bell state)
\begin{eqnarray}
\ket{p_{0}} = \ket{\beta_{11}}_{\Bcal} \ket{0}_{\Acal} = [(1/\sqrt{2})\left(\ket{01} - \ket{10}\right)_{\Bcal} ] \ket{0}_{\Acal} \text{ , }
\end{eqnarray}
Since $\hat{\Pcal}\ket{p_{0}}\!=\!0$, we then satisfy all the requirements established by the energy trapping condition. 
In addition, if we have the QB cell prepared in the state $\ket{\beta_{11}}_{\Bcal} $, then the initial energy available for the QB is $\Ecal_{0}\!=\!2\hbar \omega$. This is exactly the amount of energy required to 
offer maximum energy to the CH. Remarkably, it is possible to show that no other state, within the four Bell pairs, is capable of satisfying the trapping condition 
(see Appendix~\ref{ApC}). Therefore, we could 
start some energy flux by suitably preparing the system at these other Bell states. In fact, by preparing the system in a state $\ket{\psi_{nm}(0)}\!=\!\ket{\beta_{nm}}_{\Bcal} \ket{0}_{\Acal}$, 
where $\ket{\beta_{nm}}_{\Bcal}\!=\!\left( \ket{0n} + (-1)^{m}\ket{1\bar{n}} \right)_{\Bcal}/\sqrt{2}$ are the Bell states ($\bar{n} = 1-n$), it follows that the initial amount of energy 
in  the QB will be $\Ecal_{0}\!=\!2\hbar \omega$ for any $n,m$. From Eq.~(\ref{Ct}), the instantaneous transferred energy to the CH is
\begin{eqnarray}
C_{nm}(t) = \Ecal_{0} g_{nm} \sin^2( 2\sqrt{2} Jt) \text{ , }
\end{eqnarray}
where we have defined $g_{00}\!=\!g_{01}\!=\!1/2$, $g_{10}\!=\!1$, and $g_{11}\!=\!0$ (see Appendix~\ref{ApD}). 
Indeed, notice that the singlet state $\ket{\beta_{11}}_{\Bcal}$ retains the energy in the QB. On the other hand, the Bell states $\ket{\beta_{0m}}_{\Bcal}$ ($m=0,1$) 
allow for the transfer of $\Ecal_{0}/2$ in the discharging process, while state $\ket{\beta_{10}}_{\Bcal}$ promote a maximum energy transfer $\Ecal_{0}$, with full discharging 
time given by $\tau_{\text{d}}\!=\!\pi/(4J\sqrt{2})$. 
The advantage of using entanglement becomes evident when we consider the performance of alternative initial states.  
Indeed, as a first alternative, let us consider that the initial battery state is a factorized state, namely, 
$\ket{\psi_{\text{fc}}(0)}_{\Bcal}\!=\!\ket{\phi_{1}}_{\Bcal_{1}}  \ket{\phi_{2}}_{\Bcal_{2}}$, for arbitrary $\ket{\phi_{1}}$ and $\ket{\phi_{2}}$. 
Then, the maximum amount of energy $\Ecal_{0}$ is transferred from the QB to the CH if and only if $\ket{\phi_{1}}\!=\!\ket{\phi_{2}}\!=\!\ket{1}$, with full discharging time $\tau_{\text{fc}} = \tau_\text{d}$.  
(see Appendix~\ref{ApE}). 
This result shows that, even though a factorized initial battery state is able to provide the same EC as the Bell QB, the initial energy required by the factorized state is higher than the energy we can transfer to the CH. 
More specifically, the initial energy $\Ecal_{0}^{\text{fc}}$ associated with $\ket{\psi_{\text{fc}}(0)}_{\Bcal}\!=\!\ket{{1}}_{\Bcal_{1}}  \ket{{1}}_{\Bcal_{2}}$ is $\Ecal_{0}^{\text{fc}} = 2\Ecal_{0}$. 
Therefore, if we provide the same amount of energy $\Ecal_{0}$, the Bell QB will be more powerful than the factorized QB. As a second alternative to a Bell QB, we could also consider 
a single-particle QB, where the initial available energy is stored in a single two-level system. Again, a Bell QB will show better performance, now with respect to the transfer time. In fact, if we use single particles to 
transfer an amount of energy $\Ecal_{0}$, the transfer time is $\tau^{\text{sp}}_{\text{d}}\!=\!\pi/(4J)$, so that entangled initial states provide a gain of $\sqrt{2}$ in the EC.

\begin{figure}[t!]
	\centering
	\includegraphics[width=7.0cm]{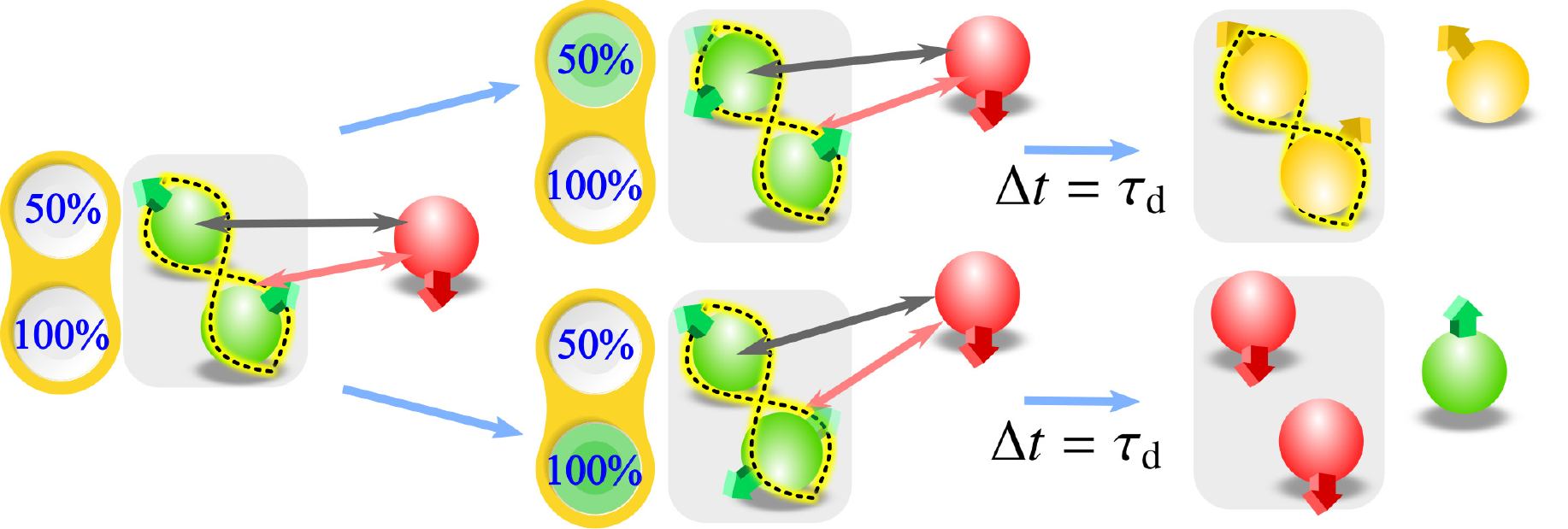}
	\caption{Schematic representation of a Bell QB cell coupled to a CH. A suitable choice of the amount of energy to be transferred can be performed 
		after a time interval $\Delta t = \tau_{\text{d}}$.}\label{FigScheme}
\end{figure}

As previously discussed and illustrated in Fig.~\ref{FigScheme},  
the Bell QB presents an inherent blocking performance by storing energy at the state $\ket{\beta_{11}}_{\Bcal}$. However, if we admit a local unitary 
operator to lead the battery state from $\ket{\beta_{11}}_{\Bcal}$ to $\ket{\beta_{10}}_{\Bcal}$, then a maximum energy $\Ecal_{0}$ will flow from the QB to the CH. 
Moreover, by using the same mechanism, we can control the portion of energy transferred to the CH. In fact, if the operation leads the battery from $\ket{\beta_{11}}_{\Bcal}$ to $\ket{\beta_{0m}}_{\Bcal}$, 
then $50\%$ of the available energy gets unblocked. Such a mechanism can be obtained by an internal agent in the QB and it does not promote 
any change of the energy in the battery, since the available energy is the same for any Bell state $\ket{\beta_{nm}}_{\Bcal}$. An activation mechanism to transfer $50\%$ of 
the energy is obtained through a bit-flip gate (Pauli operator $\sigma_{x}$) acting on one of the qubits in the QB, while the unblock of full charge is achieved by a 
phase-shift gate (Pauli operator $\sigma_{z}$) on one of the qubits in the cell (see Appendix~\ref{ApD}). It is important to mention that the transfer 
mechanism does not depend on the qubit used to activate it. This control on the energy release, with a specific ``button" designed for each situation, is illustrated in Fig.~\ref{FigScheme}.

\subsection{Scaling up cells and robustness against decoherence}

The Bell QB can be readily scaled up. Indeed, it is possible to store an amount of energy $\Ecal_{N}\!=\!N\Ecal_{0}$ by suitably building an $N$-cell Bell QB. 
This system is able to distribute energy in multiples of $\Ecal_{0}/2$ 
to the CH by controlling the internal state of the individual cells. In fact, let us consider a CH composed by $N$ qubits initially in the state $\ket{0}^{\otimes N}$. Then, we design the QB 
so that each cell can transfer its stored energy to a single qubit of the CH. As a consequence, we can provide adjustable QBs able to transfer ``quanta" of energy 
given by $\Ecal_{q}\!=\!\Ecal_{0}/2$, where we activate a number $M \leq N$ of cells of the battery. Hence, any amount of energy between $\Ecal_{q}$ and $M\Ecal_{0}$ 
can be transferred from the QB to the CH in portions of energy $\Ecal_{q}$. 

Concerning its experimental realization, 
the requirement of two-qubit XY interactions in a Bell QB allows for its implementation in currently available 
solid-state technology as, for instance, superconducting quantum circuits~\cite{Clarke:08,circQED1,wallraff2004}, quantum dots and cavity QED~\cite{Imamoglu:99}, atoms in a 
one-dimensional trap~\cite{Volosniev:14,Volosniev:15,Levinsen:15}, tunable microscopic optical traps~\cite{Murmann:15-a,Murmann:15-b}, and among others~\cite{Pellegrino:14,majer2007}. 
Moreover, it is relevant to mention the expected robustness of Bell QBs against decohering effects. Indeed, we can take advantage of Bell QBs in some relevant physical 
scenarios, where the dominant decohering effect is due to \textit{collective dephasing} processes. Indeed, collective dephasing is a non-unitary phenomenon associated 
with the dynamics of $N$ non-interacting particles sharing the same environment~\cite{Carnio:16}, which occurs due to fluctuations of the field that acts on each 
particle~\cite{Gross:10,Schindler:13,Hu:18,Hu:19-a} or due to dipole-dipole interactions~\cite{Prasanna:18}. Under these non-unitary effects, there is an intrinsic 
\textit{decoherence-free subspace} spanned by the states $\ket{10}$ and $\ket{01}$, leading to the complete robustness of the Bell state $\ket{\beta_{11}}_{\Bcal}$~\cite{Carnio:16,Lidar:03} 
(see also a recent loss-free scheme for excitonic QBs based on a symmetry-protected dark state~\cite{Liu:19}). 
Naturally, non-unitary effects may appear when we couple the CH to the QB, but it can be smoothened by adjusting $J$ so that $\tau_{\text{d}} \ll \tau_{\text{dp}}$, where 
$\tau_{\text{dp}}$ is the relaxation time scale.

\section{Adiabatic model for stable QBs}

In the Bell QB scheme previously introduced, we have a phenomenon known as \textit{spontaneous discharge}~\cite{Santos:19-a}. 
This is a typical consequence of the unitary dynamics, where the energy oscillates between the QB and the CH after the QB starts its energy distribution. In turn, there will be energy revivals  
in the QB for specific times, which prevents a stable battery discharge. In order to circumvent this problem, Ref.~\cite{Santos:19-a} 
has shown a specific situation where a stable charging process can be achieved by using the adiabatic dynamics specifically derived for a three-level system. 
Here, inspired by Eq.~(\ref{potadi}), we can provide a general approach to stabilize the charging/discharging processes of QBs. In order to show that the adiabatic dynamics provides a stable process for both charge and discharge of the QB, we will analyze them separately. 

The charging process is directly achieved by adiabatic engineering state of a Bell state~\cite{Unanyan:01}, where the battery system is adiabatically driven by a time dependent Hamiltonian $H_\Bcal(t)$ from initial state $\ket{00}_{\Bcal}$ to the final state $\ket{\beta_{11}}_{\Bcal}$. Since the protocol starts from a single state of $H_\Bcal(0)$, Eq.~\eqref{potadi} guarantees that the charging process will be stable. 
In the discharging process, quantum control is provided by a time-dependent Hamiltonian $H(t)$. We assume that $H(t)$ can be turned on and off. 
As we turn it on, it is also able to ensure stability in the discharge process. 
As in the case of the Bell QB, we begin by taking the initial state of the whole system as $\ket{\psi(0)}\!=\!\ket{\beta_{11}}_{\Bcal} \ket{0}_{\Acal}$. 
Then, let us consider a suitable three-qubit time-dependent Hamiltonian, which reads
\begin{equation}
H(t) = [1-f(t)] H_{\text{i}} + [1-f(t)]f(t) H_{\text{m}} + f(t) H_{\text{f}} \text{ , }
\end{equation}
where $H_{\text{i}}\!=\! \hbar J (\sigma_{x}^{\Bcal_1}\sigma_{x}^{\Bcal_2} + \sigma_{y}^{\Bcal_1}\sigma_{y}^{\Bcal_2})$, $H_{\text{f}}\!=\! \hbar J ( \sigma_{z}^{\Bcal_1}\sigma_{z}^{\Acal} + \sigma_{z}^{\Bcal_2}\sigma_{z}^{\Acal} )$, and $H_{\text{m}}\!=\! \hbar J ( \sigma_{x}^{\Bcal_1}\sigma_{x}^{\Bcal_2} + \sigma_{y}^{\Bcal_1}\sigma_{y}^{\Bcal_2} + \sigma_{x}^{\Bcal_2}\sigma_{x}^{\Acal} + \sigma_{y}^{\Bcal_2}\sigma_{y}^{\Acal} )$ are the initial, final and middle Hamiltonians, respectively. The Hamiltonian $H(t)$ fulfills the following required properties: (i) $\ket{\psi(0)}$ is ground state of $H_{\text{i}}$; 
(ii) the final desired state $\ket{00}_{\Bcal} \ket{1}_{\Acal}$ is the ground state of $H_{\text{f}}$; (iii) there are no level crossings, which is assured by $H_{\text{m}}$.
This allows for the quantum evolution towards the target state through the adiabatic dynamics.

\begin{figure}[t!]
	\centering
	\includegraphics[width=6.3cm]{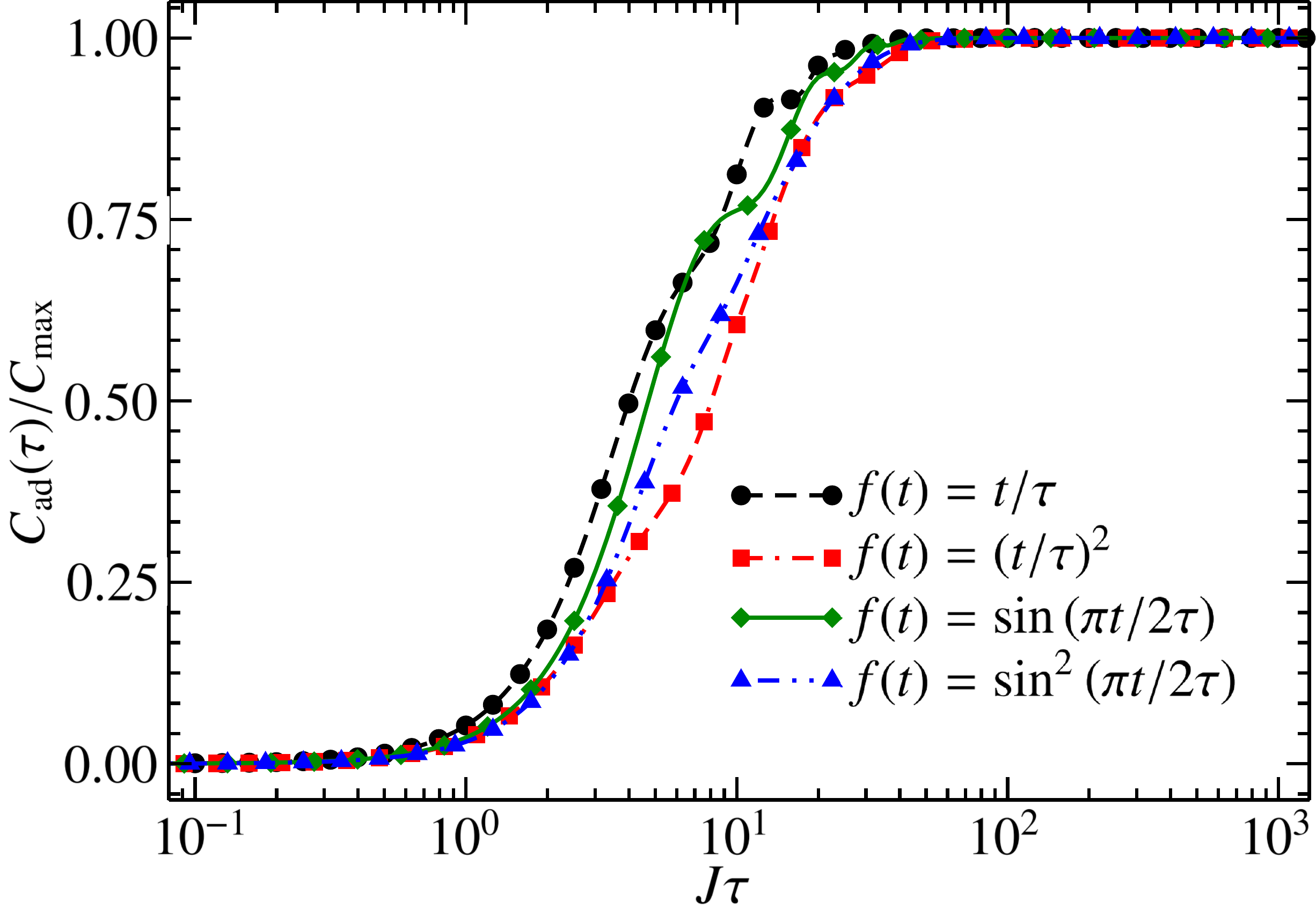}
	\vspace{-0.2cm}
	\caption{Amount of energy, as multiple of $C_{\text{max}}$, transferred from the QB to the CH 
		through an adiabatic dynamics as function of the parameter $J\tau$.}\label{FigAdiabatic}
\end{figure}

Due the double degenerate spectrum of the ground state of $H_{\text{fin}}$, it remains to prove that the final state $\ket{00}_{\Bcal} \ket{1}_{\Acal}$ is achieved due to the symmetries of $H(t)$. 
Indeed, the two independent ground states of $H_{\text{fin}}$ are $\ket{\psi_{0}}\!=\!\ket{11}_{\Bcal} \ket{0}_{\Acal}$ and $\ket{\psi_{1}}\!=\!\ket{00}_{\Bcal} \ket{1}_{\Acal}$. Then, by defining the 
parity operator $\Pi_{z}\!=\!\sigma_{z}^{\Bcal_1}\sigma_{z}^{\Bcal_2}\sigma_{z}^{\Acal}$, we can verify that $[H(t),\Pi_{z}]\!=\!0$ for all $t$. Therefore, $H(t)$ preserves the parity of the state 
throughout the evolution. Since we start the evolution at the state $\ket{\psi(0)}$, which has parity eigenvalue $-1$, the system will evolve to subsequent states with the same parity. Hence, transitions to 
$\ket{\psi_{0}}$ at the end of the evolution are forbidden, so that all the energy of the QB can be adiabatically transferred to the CH, achieving the final state $\ket{\psi_{1}}$. 
This is illustrated in Fig.~\ref{FigAdiabatic}, which shows that this transfer is indeed stable  
for several different choices of the interpolation function $f(t)$. Notice then that, in agreement with Eq.~(\ref{potadi}), the QB discharge process is stable and robust against variations of the interpolation scheme.

\section{Conclusion}

A general design of a QB has been proposed, which originated from the EC operator associated with a suitable interaction Hamiltonian. The QB proposal provides a scalable energy source fueled by Bell pairs, with the stability achieved through a local Hamiltonian.  
We have then shown that the EC operator is the source of two main applications: (i) a trapping energy 
mechanism based on a common eigenstate between the EC operator and the interaction Hamiltonian, in which the battery can indefinitely retain 
its energy even if it is coupled to the CH; (ii) an asymptotically stable discharge mechanism, which is achieved by adiabatic evolutions leading to 
vanishing EC. The first application highlights the strength of entanglement for QBs, which is advantageous both for the fine grained control of the energy discharge 
and for the gain of $\sqrt{2}$ in the energy transfer rate for each individual cell in the QB. The second application solves in general the stability of the energy transfer process, 
with no backflow of energy from the CH to the QB. More specifically, we introduced a piecewise time-dependent Hamiltonian to stabilize the energy discharging process through 
the adiabatic dynamics. In both cases, the scaling of the QB can be achieved by adding Bell pairs to the QB, with each Bell pair providing a controllable mechanism for full and 
stable charge of a qubit in the CH. Experimental implementations may inherit hardware designs originally proposed for adiabatic quantum computing, available with current technology.

\begin{acknowledgments}
A.C.S. is supported by Conselho Nacional de Desenvolvimento Cient\'{\i}fico e Tecnol\'ogico (CNPq-Brazil). 
M.S.S. is supported by CNPq-Brazil (No. 303070/2016-1) and Funda\c{c}\~ao Carlos Chagas Filho de Amparo \`a Pesquisa do Estado do Rio de Janeiro (FAPERJ) (No. 203036/2016). 
The authors also acknowledge financial support in part by the Coordena\c{c}\~ao de Aperfei\c{c}oamento de Pessoal de N\'{\i}vel Superior - Brasil (CAPES) (Finance Code 001) and by the Brazilian
National Institute for Science and Technology of Quantum Information [CNPq INCT-IQ (465469/2014-0)]. 
\end{acknowledgments}

\appendix

\section{The energy current observable} \label{ApA}

Consider a bipartite system in the Hilbert space $\Bcal \otimes \Acal$, where $\Bcal$ is associated with the quantum battery (QB) and $\Acal$ 
denotes an auxiliary system associated with a consumption hub (CH). 
The system dynamics is governed by the Hamiltonian
\begin{equation}
H(t) = H_{0} + H_{\text{C}}(t) \text{ , } \label{Ht}
\end{equation}
where $H_{0}$ is the natural Hamiltonian system which defines the energy states of the system and $H_{\text{C}}(t)$ 
is the interaction Hamiltonian between the QB and the CH. 
Without loss of generality, we take
\begin{equation}
H_{0} = \sum\nolimits_{n=\{\Ccal,\Bcal\}} H_{0}^{(n)} \text{ , }
\end{equation}
where $H_{0}^{\Bcal}$ and $H_{0}^{\Acal}$ are the bare Hamiltonians of the QB and CH, respectively. 
We then define the energy stored in battery as
\begin{equation}
C(t) = \tr{H_{0}^{\Acal}\rho(t)} - E^{\Acal}_{\text{emp}} \text{ , }
\end{equation}
where $\rho$ is the density matrix for a pure state of the composite system and $E^{\Acal}_{\text{emp}}$ is the energy of the passive auxiliary state (empty charge state). 
Here, we have $E^{\Acal}_{\text{emp}}\!=\!\trs{ H^{\Acal}_{0}\rho^{\Acal}_{\text{gs}} }$, with $\rho^{\Acal}_{\text{gs}}$ denoting the ground state of $H^{\Acal}_{0}$. 
By considering the Schr\"odinger equation in the interaction picture, we have
\begin{equation}
\dot{\rho}_{\text{int}}(t) = 1/(i\hbar) [H_{\text{int}},\rho_{\text{int}}(t)] \text{ , } \label{VonNeuInt}
\end{equation}
where the dot symbol denotes time derivative, $H_{\text{int}}\!=\!\Zcal^{\dagger}(t) H_{\text{C}} \Zcal(t)$, and 
$\rho_{\text{int}}(t)\!=\!\Zcal^{\dagger}(t)\rho(t)\Zcal(t)$, with $\Zcal^{\dagger}(t)\!=\!e^{iH_{0}t/\hbar}$. 
It is important to mention that the relevant quantities to be computed here are independent of the frame 
used to study the dynamics. In fact, by computing $C(t)$  
in the interaction picture, we get
\begin{align}
C_{\text{int}}(t) &= \tr{H_{0}^{\Acal}\rho_{\text{int}}(t)} - E^{\Acal}_{\text{emp}}  \nonumber \\
&= \tr{H_{0}^{\Acal}\Zcal^{\dagger}(t)\rho(t)\Zcal(t)} - E^{\Acal}_{\text{emp}}  \nonumber \\
&= \tr{\Zcal^{\dagger}(t)H_{0}^{\Acal}\rho(t)\Zcal(t)} - E^{\Acal}_{\text{emp}}  \nonumber \\
&= \tr{H_{0}^{\Acal}\rho(t)\Zcal(t)\Zcal^{\dagger}(t)} - E^{\Acal}_{\text{emp}} 
\nonumber \\
&= \tr{H_{0}^{\Acal}\rho(t)} - E^{\Acal}_{\text{emp}}  = C(t) \text{ . }
\end{align}
As a first result from the above discussion, we can define a Hermitian operator $\hat{\Pcal}(t)$ associated with the instantaneous battery charging. 
By defining the instantaneous EC as $\Pcal(t)\!=\!\dot{C}(t)$, we get
\begin{equation}
\Pcal(t) = \dot{C}(t) = \tr{H_{0}^{\Acal}\dot{\rho}(t)} = \frac{1}{i\hbar}\tr{H_{0}^{\Acal}[H(t),\rho(t)]} \text{ , }
\end{equation}
where we used Schr\"odinger equation for $\rho(t)$. Thus, we find
\begin{align}
\Pcal(t) &= \frac{1}{i\hbar}\tr{H_{0}^{\Acal}\left[H(t)\rho(t)-\rho(t)H(t)\right]} \nonumber \\
&= \frac{1}{i\hbar}\tr{H_{0}^{\Acal}H(t)\rho(t)-H(t)H_{0}^{\Acal}\rho(t)} \nonumber \\
&= \frac{1}{i\hbar}\tr{ [H_{0}^{\Acal},H(t)]\rho(t)}= \frac{1}{i\hbar}\tr{ [H_{0}^{\Acal},H_{\text{C}}(t)]\rho(t)} \text{ , }
\end{align}
where we used Eq.~\eqref{Ht}. Then, by defining
\begin{equation}
\hat{\Pcal}(t) = \frac{1}{i\hbar}[H_{0}^{\Acal},H_{\text{C}}(t)] \text{ , } \label{Eq}
\end{equation}
we conclude that
\begin{equation}
\Pcal(t) = \tr{ \hat{\Pcal}(t)\rho(t)} \text{ . }
\end{equation}
In the interaction picture, we can use the Eq.~\eqref{VonNeuInt} to find that 
$P_{\text{int}}(t)\!=\!\tr{ \hat{\Pcal}_{\text{int}}(t)\rho_{\text{int}}(t)}\!=\!\Pcal(t)$, 
where $P_{\text{int}}(t)$ is the EC in interaction picture, which reads $\hat{\Pcal}_{\text{int}}(t)\!=\![H_{0}^{(\Acal)},H_{\text{int}}]/(i\hbar)$.

\section{Adiabatic QBs are stable} \label{ApB}

As an application of the definition of the EC operator, let us show here that adiabatic QBs are stable. 
We say that the system QB-CH undergoes an adiabatic transfer process when the composite quantum state 
$\ket{\psi(t)} \in \Bcal \otimes \Acal$ evolves under adiabatic dynamics for the Hamiltonian $H(t)\!=\!H_{0} + H_{\text{C}}(t)$. 
Let the system be initialized in the state
\begin{equation}
\ket{\psi(0)} = \sum_{n} c_{n} \ket{E_{n}(0)} \text{ , }
\end{equation}
where $\{\ket{E_{n}(0)}\}$ is the set of eigenvectors of the Hamiltonian $H(t)$ at $t\!=\!0$. Therefore, in the adiabatic regime, we have
\begin{equation}
\ket{\psi_{\text{ad}}(t)} = \sum_{n} c_{n} e^{i\theta^{\text{ad}}_{n}(t)}\ket{E_{n}(t)} \text{ , }
\end{equation}
where $\theta^{\text{ad}}_{n}(t)$ comprises both the dynamic and geometric adiabatic phases associated with the eigenstate  $\ket{E_{n}(t)}$. 
Therefore
\begin{align}
\Pcal_{\text{ad}}(t) &= \tr{ \hat{\Pcal}(t)\rho_{\text{ad}}(t)} = \frac{1}{i\hbar}\tr{[H_{0}^{\Acal},H_{\text{C}}(t)]\rho_{\text{ad}}(t)} \nonumber \\
&= \frac{1}{i\hbar} \sum_{n,m} c_{n} c^{\ast}_{m} e^{i\Delta^{\text{ad}}_{nm}(t)} \bra{E_{n}(t)} [H_{0}^{\Acal},H(t)] \ket{E_{m}(t)}  \nonumber \text{ , }
\end{align}
with $\Delta^{\text{ad}}_{nm}(t)\!=\!\theta^{\text{ad}}_{n}(t) - \theta^{\text{ad}}_{m}(t)$ and where we have used 
\begin{equation}
[H_{0}^{\Acal},H(t)] = [H_{0}^{\Acal},H_{0}^{\Acal} + H_{0}^{\Bcal} + H_{\text{C}}(t)] = [H_{0}^{\Acal}, H_{\text{C}}(t)] \text{ .}
\end{equation}
Thus, we get
\begin{equation}
\Pcal_{\text{ad}}(t) = \frac{1}{i\hbar} \sum_{n,m} c_{n} c^{\ast}_{m} e^{i\Delta^{\text{ad}}_{nm}(t)} \Delta \tilde{E}_{mn}
\text{ , } 
\end{equation}
where $\Delta \tilde{E}_{mn}\!=\!  \left[E_{m}(t)-E_{n}(t)\right]\bra{E_{n}(t)}H_{0}^{\Acal}\ket{E_{m}(t)} $. Now, let us consider the case where our system starts in a single eigenstate of the Hamiltonian $H(0)$, for example the $k$-th eigenstate of $H(0)$. By doing that, we have $c_{n}\!=\!\delta_{nk}$ and the above equation becomes
\begin{equation}
\Pcal_{\text{ad}}(t) = \frac{1}{i\hbar} \sum_{n,m} \delta_{nk} \delta_{mk} e^{i\Delta^{\text{ad}}_{nm}(t)}\Delta \tilde{E}_{mn} = \frac{e^{i\Delta^{\text{ad}}_{kk}(t)} }{i\hbar}\Delta \tilde{E}_{kk} = 0\text{ . } 
\end{equation}
Therefore, since the stability condition for an interval $[t_{1},t_{2}]$ can be mathematically written as $\Pcal(t)\!=\!0$ for any $t \in [t_{1},t_{2}]$, this result allows us to conclude that adiabatic dynamics is a strategy to get a stable charging process in quantum batteries because $\Pcal_{\text{ad}}(t\geq \tau_{\text{ad}})\!=\!0$, where $\tau_{\text{ad}}$ is the required time to achieve the adiabatic regime. The same calculation as be done for the situation where we have a degenerate eigenspace of the Hamiltonian, where get the same result.

\section{Bell singlet state and energy transfer protection} \label{ApC}

Here we want to show that the initial state able to block the energy transfer from the QB to the CH is unique and 
given by the Bell singlet state $\ket{\beta_{11}} = (1/\sqrt{2}) (|01\rangle - |10\rangle)$. 
To this end, let us start by considering a general initial density operator $\rho$ with matrix elements $\rho_{ij}$. 
The initial state $\rho$ is expected to satisfy the conditions: 
\begin{itemize}
	\item[(Ca)] The available energy is $\Ecal_{0} = 2\hbar \omega$.
	\item[(Cb)] The instantaneous EC is $\Pcal(t)= 0, \forall t$.
\end{itemize}
By computing the available energy for $\rho$, we get
\begin{align}
\Ecal_{0} = \hbar \omega(2 + \rho_{11}- \rho_{44} ) \text{ , }
\end{align}
so that the condition (Ca) imposes $\rho_{11}\!=\!\rho_{44}$. Now, by computing the EC, we find
\begin{align}
\Pcal (t) &= 2\sqrt{2} J\omega \hbar \sin \left( 4\sqrt{2} Jt \right) \left( 2\rho_{11} + \rho_{22} + \rho_{33} \right) \nonumber \\
&= 4 \sqrt{2} J \text{Re}[\rho_{23}] \omega \hbar \sin \left( 4\sqrt{2} Jt \right) \text{ . } \label{Power-App}
\end{align}
From Eq.~\eqref{Power-App}, we note that there are a number of elements of $\rho$ that do not contribute to EC and, for simplicity, we just take them as zero. 
In particular, we adopt $\rho_{23}\in \Rmath$. 
Condition (Cb) then imposes that $2\rho_{11} + \rho_{22} + \rho_{33} +2 \rho_{23}\!=\!0$, 
Summarizing, by taking into account conditions (Ca) and (Cb), as well as the requirements of probability conservation and positivity of $\rho$, we obtain the set of constraints
	\begin{align}
	\rho_{11}  &= \rho_{44}\text{ , } \label{Cond1} \\ 
	2\rho_{11} + \rho_{22} + \rho_{33} +2 \rho_{23} &= 0 \text{ , } \label{Cond2}\\
	\sum_{n=1}^{4} \rho_{nn} &= 1 \text{ , } \label{Cond3}\\
	0 \le 1 - 4 \left( \rho_{22}\rho_{33} - \rho_{23}^2\right) &\le 1 \text{ . } \label{Cond4}
	\end{align}

From the constraints~\eqref{Cond1},~\eqref{Cond2} and~\eqref{Cond3}, we obtain
\begin{align}
\rho_{22} &= 1 - 2\rho_{11} - \rho_{33} \,\,\,\text{ and } \,\,\, \rho_{23} = -1/2 \text{ . }
\label{C4}
\end{align}
Therefore, it follows that the density matrix can be written as
\begin{align}
\rho = \begin{bmatrix}
\rho_{11} & 0 & 0 & 0 \\
0 & 1 - 2\rho_{11} - \rho_{33} & -1/2 & 0 \\
0 & -1/2 & \rho_{33} & 0 \\
0 & 0 & 0 & \rho_{11}
\end{bmatrix} \text{ . } \label{ApRhoMatrix}
\end{align}
In addition, by using \eqref{Cond4}, we obtain the inequality
\begin{align}
\rho_{22} \rho_{33} - \rho_{23}^2 &\geq 0 \text{ , } \label{EqApRho33-1} 
\end{align}
From \eqref{C4} and \eqref{EqApRho33-1}, we find
\begin{align}
- \rho_{33}^2 + (1-2\rho_{11})\rho_{33} &\geq  1/4  \text{ . }
\label{C8}
\end{align}
By maximizing the function $f(\rho_{33}) = - \rho_{33}^2 + (1-2\rho_{11})\rho_{33} $, it follows that the maximum point $f_{\text{max}}$ of  $f(\rho_{33})$ is 
provided by $f_{\text{max}} = (1-2\rho_{11})/4$. However, \eqref{C8} requires that $f(\rho_{33}) \ge 1/4$. This yields $\rho_{11} = 0$. 
Moreover, $f_{\text{max}}$ is obtained if and only if $\rho_{33} = (1-2\rho_{11})/2$, which then implies that $\rho_{33} = 1/2$.  
In conclusion, the admissible initial density matrix that is able to retain energy in the QB is essentially unique and reads
\begin{align}
\rho = \ket{\beta_{11}} \bra{\beta_{11}} \text{ . }
\end{align}

\section{Energy transfer from Bell QBs} \label{ApD}

Let us consider the initial state for the CH as the empty state $\ket{0}_{\Acal}$ and that the QB is prepared in a Bell state 
\begin{align}
\ket{\beta_{nm}}_{\Bcal} = \frac{1}{\sqrt{2}}\left( \ket{0n} + (-1)^{m}\ket{1\bar{n}} \right)_{\Bcal} \text{ . }
\end{align}
The system dynamics driven by the Hamiltonian 
\begin{eqnarray}
H_{\text{int}} = \hbar J \sum\nolimits_{n=1}^{2}\left( \sigma_{x}^{\Bcal_n}\sigma_{x}^{\Acal} + \sigma_{y}^{\Bcal_n}\sigma_{y}^{\Acal} \right) \text{ . } \label{HChargeAp}
\end{eqnarray}
Then, we can write the instantaneous evolved state as
\begin{align}
\ket{\psi_{nm}(t)} = \exp\left( -\frac{i}{\hbar} H_{\text{int}} t \right) \ket{\beta_{nm}}_{\Bcal} \ket{0}_{\Acal}\text{ . }
\end{align}
By computing the instantaneous amount of energy transferred to the CH, we obtain
\begin{align}
C_{nm}(t)=\bra{\psi_{nm}(t)}H_{0}^{\Acal}\ket{\psi_{nm}(t)} - E(0) \text{ , }
\end{align}
where $H_{0}^{\Acal}\!=\! \1_{\Bcal} \otimes (\hbar \omega \sigma_{z})_{\Acal}$ and $E(0)\!=\!_{\Acal}\bra{0} (\hbar \omega \sigma_{z})_{\Acal} \ket{0}_{\Acal} = -\hbar\omega$ is the initial energy in the CH, 
which is then fully discharged. 
Therefore
\begin{subequations}\label{ApEqCharge}
	\begin{align}
	C_{00}(t) &= C_{01}(t) = \frac{1}{2}\hbar \omega \sin^2 \left( 2\sqrt{2} Jt \right) \text{ , }\\
	C_{10}(t) &= \hbar \omega \sin^2 \left( 2\sqrt{2} Jt \right) \text{ , } \\
	C_{11}(t) &= 0 \text{ , }
	\end{align}
\end{subequations}
so that the above equations can be rewritten as
\begin{eqnarray}
C_{nm}(t) = \Ecal_{0} g_{nm} \sin^2 \left( 2\sqrt{2} Jt \right) \text{ , }
\end{eqnarray}
where we have defined $g_{00}\!=\!g_{01}\!=\!1/2$, $g_{10}\!=\!1$, and $g_{11}\!=\!0$. 
From above equations, in order to control the amount of energy transferred to CH, we can start the QB in the state 
$\ket{\beta_{11}}_{\Bcal}$, so that no energy is transferred to the CH. Now, in order to transfer $50\%$ of the available energy 
we need to change the Bell state from $\ket{\beta_{11}}_{\Bcal}$ to $\ket{\beta_{0m}}_{\Bcal}$. This can be achieved by implementing the operation 
$\sigma_{x}$ in the first qubit in the QB. In fact, by implementing the operation $O^{(2)}_x = \1^{\Bcal_{1}} \otimes \sigma_{x}^{\Bcal_{2}}$ we have (up to a global phase)
\begin{align}
\ket{\beta_{11}}_{\Bcal} \rightarrow O^{(2)}_x \ket{\beta_{11}}_{\Bcal} = \ket{\beta_{01}}_{\Bcal} \text{ , }
\end{align}
so that after this operation the energy transferred to CH is given by $C_{01}(t)$. 
The same result is obtained if we implement the operation $O^{(1)}_x = \sigma_{x}^{\Bcal_{2}} \otimes \1^{\Bcal_{1}}$, because
\begin{align}
\ket{\beta_{11}}_{\Bcal} \rightarrow O^{(1)}_x \ket{\beta_{11}}_{\Bcal} = \ket{\beta_{01}}_{\Bcal} \text{ . }
\end{align}

In both cases, the maximum energy transferred is $\Ecal_{0}/2$, namely, $50\%$ of the available energy in the battery. 
On the other hand, from Eqs.~\eqref{ApEqCharge} we can transfer $100\%$ of the available energy if we change the initial state 
$\ket{\beta_{11}}_{\Bcal}$ to $\ket{\beta_{10}}_{\Bcal}$. This can be performed through one of the two phase-shift operations given by

\begin{align}
O^{(1)}_z = \sigma_{z}^{\Bcal_{1}} \otimes \1^{\Bcal_{2}} \text{ , \ or \ } O^{(2)}_z = \1^{\Bcal_{1}} \otimes \sigma_{z}^{\Bcal_{2}} \text{ . }
\end{align}

\begin{figure}[b!]
	\centering
	\includegraphics[width=5.4cm]{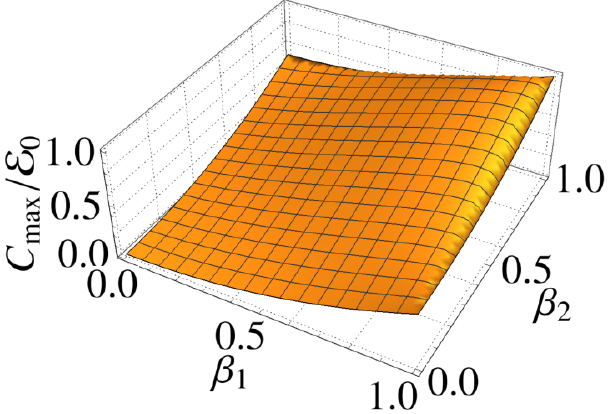}
	\caption{Maximum value of the charge transferred to the CH for separable states as function of the parameters $\beta_{1}$ and $\beta_{2}$.}
	\label{3dGraph}
\end{figure}

Therefore, it follows that the energy transfer mechanism is independent of the qubit used to activate the process.

\section{Performance of QBs based on separable states} \label{ApE}

Let us consider the discharging process as performed by Bell QBs, but taking now a separable state for the energy transfer process. Then, the Hamiltonian is given by Eq.~\eqref{HChargeAp}, where the initial battery state is assumed to be a separable state $\ket{\psi_{\text{fc}}(0)}_{\Bcal}\!=\!\ket{\phi_{1}}_{\Bcal_{1}}  \ket{\phi_{2}}_{\Bcal_{2}}$. Here, we take 
$\ket{\phi_{n}}\!=\!\alpha_{n}\ket{0} + \beta_{n}e^{i\theta_{n}}\ket{1}$, with arbitrary amplitudes $0\leq \alpha_n \leq 1$ and $0 \leq \beta_n \leq 1$.
Then, by letting the system evolve and computing the instantaneous energy, we find
\begin{align}
\frac{C_{\text{fc}}(t)}{\hbar\omega} &= 2 \beta_{1} \beta_{2} \tilde{\beta}_{1}\tilde{\beta}_{2} \cos (\theta_{1} - \theta_{2}) \sin^2(2 \sqrt{2} Jt)\nonumber \\ &+ (\beta_{1}^2 + \beta_{2}^2) \sin^2(2 \sqrt{2} Jt) \text{ , }
\end{align}
where $\tilde{\beta}_{n}^2\!=\!1 - \beta_{n}^2$, with the subscript ``fc" denoting the energy for ``factorized" states and where $2\hbar\omega$ is the maximum charge to be transferred to the CH. Therefore, by studying the maxima and minima of the above function, we obtain 
the critical values for $C_{\text{fc}}(t)$ as $t_{\text{c},n}\!=\!n\pi/(4\sqrt{2}J)$. Since the initial value of $C_{\text{fc}}(0)$ is a minimum of $C_{\text{fc}}(t)$, 
then the first maximum happens when $\tau\!=\!t_{\text{c},1}\!=\!\pi/(4\sqrt{2}J)$, where the charge reads
\begin{align}
\frac{C_{\text{max}}}{\hbar\omega} &= 2 \beta_{1} \beta_{2} \tilde{\beta}_{1}\tilde{\beta}_{2} \cos (\theta_{1} - \theta_{2}) + \beta_{1}^2 + \beta_{2}^2 \text{ . }
\end{align}
Since we need to get $C_{\text{max}}\!=\!\Ecal_{0}\!=\!2\hbar\omega$, the first optimization can be done in the parameters $\theta_{n}$, where we need to have $\theta_{1} - \theta_{2}\!=\!2n\pi$, so that
\begin{align}
\frac{C_{\text{max}}}{\Ecal_{0}} &= \beta_{1} \beta_{2} \sqrt{\left( 1 - \beta_{1}^2 \right) \left(1 - \beta_{2}^2\right)} + \frac{\beta_{1}^2 + \beta_{2}^2}{2} \text{ . }
\end{align}
In Fig.~\ref{3dGraph} we show the plot of $C_{\text{max}}/\Ecal_{0}$ as a function of the parameters $\beta_{1}$ and $\beta_{2}$, where we can see that the maximum charge is obtained when $\beta_{1}\!=\!\beta_{2}\!=\!1$. 
Therefore, the maximum amount of energy $\Ecal_{0}$ is transferred from the QB to the CH if and only if $\ket{\phi_{1}}\!=\!\ket{\phi_{2}}\!=\!\ket{1}$, where the initial available energy $\Ecal_{0}^{\text{fc}}$ is $\Ecal_{0}^{\text{fc}}\!=\!4\hbar \omega\!=\!2 \Ecal_{0}$.


%

\end{document}